\documentclass[runningheads]{llncs}
\usepackage[T1]{fontenc}
\usepackage{graphicx}
\usepackage{svg}
\usepackage{amsmath}
\usepackage{cite}
\usepackage{hyperref}
\usepackage{color, soul}
\usepackage{xcolor,url}
\usepackage{subfigure}
\usepackage{multirow}
\usepackage{amssymb}
\usepackage{amsfonts}
\graphicspath{ {./images/} }
\usepackage{flushend} % 参考文献双栏平均
\usepackage{ulem}    % 参考文献去掉下划线
\usepackage{longtable}
\usepackage{xparse}
\usepackage{booktabs}
\usepackage{subfigure}
\usepackage{subcaption}
\usepackage{caption}

\usepackage{array}
\usepackage{diagbox}
\begin{document}
\title{MCDubber: Multimodal Context-Aware Expressive Video Dubbing}
\titlerunning{Multimodal Context-Aware Expressive Video Dubbing}
\author{Yuan Zhao\and Zhenqi Jia 
% $^{\dag}$
\and Rui Liu\thanks{: Corresponding Author. 
% $^{\dag}$: Equal Contributions. 
This research was funded by the Young Scientists Fund of the National Natural Science Foundation of China (No. 62206136), Guangdong Provincial Key Laboratory of Human Digital Twin (No. 2022B121201 0004),
% the High-level Talents Introduction Project of Inner Mongolia University (No. 10000-22311201),
and the ``Inner Mongolia Science and Technology Achievement Transfer and Transformation Demonstration Zone, University Collaborative Innovation Base, and University Entrepreneurship Training Base'' Construction Project (Supercomputing Power Project) (No.21300-231510). This research of De Hu was funded by the National Natural Science Foundation of China under Grants 62361045 and 62201297. } \and De Hu\and Feilong Bao \and Guanglai Gao }
\authorrunning{Yuan Zhao \& Zhenqi Jia \&Rui Liu et al.}
\institute{Inner Mongolia University, Hohhot, China\\
\email{zy404nf@163.com, \ jiazhenqi7@163.com, \   liurui\_imu@163.com,\ cshood@imu.edu.cn,     \{csfeilong, csggl\}@imu.edu.cn}}

\maketitle              % typeset the header of the contribution

\begin{abstract}

Automatic Video Dubbing (AVD) aims to take the given script and generate speech that aligns with lip motion and prosody expressiveness. Current AVD models mainly utilize visual information of the current sentence to enhance the prosody of synthesized speech. However, it is crucial to consider whether the prosody of the generated dubbing aligns with the multimodal context, as the dubbing will be combined with the original context in the final video. This aspect has been overlooked in previous studies.
To address this issue, we propose a Multimodal Context-aware video Dubbing model, termed \textbf{MCDubber}, to convert the modeling object from a single sentence to a longer sequence with context information to ensure the consistency of the global context prosody.
MCDubber comprises three main components: 
(1) A context duration aligner aims to learn the context-aware alignment between the text and lip frames;
(2) A context prosody predictor seeks to read the global context visual sequence and predict the context-aware global energy and pitch;
(3) A context acoustic decoder ultimately predicts the global context mel-spectrogram with the assistance of adjacent ground-truth mel-spectrograms of the target sentence.
Through this process, MCDubber fully considers the influence of multimodal context on the prosody expressiveness of the current sentence when dubbing. The extracted mel-spectrogram belonging to the target sentence from the output context mel-spectrograms is the final required dubbing audio.
Extensive experiments on the Chem benchmark dataset demonstrate that our MCDubber significantly improves dubbing expressiveness compared to all advanced baselines.
The code and demos are available at \url{https://github.com/XiaoYuanJun-zy/MCDubber}.

\keywords{Automatic Video Dubbing \and Multimodal Context \and  Prosody Expressiveness.}
\end{abstract}

\section{Introduction}
Dubbing is a post-production process where professional actors re-record dialogues for videos to enhance audio quality \cite{hu2021neural}. With the rise of short videos and the growth of the film industry, the demand for video dubbing has surged. However, the process of dubbing is costly, requiring studios, professional actors, and significant recording time \cite{3D-VD}. The Automatic Video Dubbing (AVD) task aims to meet the growing demand for video dubbing by converting a given script into speech that aligns with lip motion and prosody expressiveness. Text-to-speech (TTS) technologies, such as Tacotron 1/2 \cite{tacotron1,tacotron2}, FastSpeech 1/2 \cite{fastspeech1, fastspeech2} , have been instrumental in advancing AVD. However, AVD is more challenging than TTS tasks since it requires generated speech that aligns with lip motion and prosody expressiveness. In recent years, numerous outstanding achievements have arisen in AVD.
 %(see Figure\ref{fig_avd})%
% \begin{figure}[h!]
% \centering
% \resizebox{0.9\linewidth}{!}{\includegraphics{images/AVD.png}}
% \caption{An illustration of the AVD workflow: The model takes video frames and corresponding scripts as input, and outputs speech that aligns with both lip motion and prosody expressiveness.}
% \label{fig_avd}
% \end{figure}

Current AVD methods generally focus on two main issues. 1) Duration control and audio-visual synchronization. Neural Dubber \cite{hu2021neural} and DSU-AVO \cite{DSU-AVO} employ a text-video aligner by cross-modal attention to learn the single sentence alignment between lip motion and text, aiming to address these challenges. 2) Prosody expressiveness in video dubbing. VDTTS \cite{VDTTS}, HPMDubbing \cite{HPMDubbing}, and 3D-VD \cite{3D-VD} improve the prosody expressiveness of dubbing by integrating facial visual information of current (target) sentence. These methods are sometimes limited by their insufficient consideration of the multi-modal context of the current sentence being dubbed.

Despite the progress, we note that to enhance prosody expressiveness in video dubbing, integrating multi-modal context information is a promising approach. Moreover, ensuring seamless alignment between the prosody of the generated dubbing and the multimodal context is crucial, given that the dubbing will eventually merge with the original context in the final video. Existing TTS models effectively utilize acoustic and textual contexts to enhance the prosody of generated speech \cite{masked_speech, context_speech, oplustil2020using}. In real dubbing scenarios, voice actors use scripts and silent video frames of the current sentence, while considering the previous and following videos, to ensure the dubbing aligns with the multimodal context.
%加参考文献，实在找不到参考文献
Therefore, how to explore the integration of multi-modal context information into the AVD task to further improve the context-aware prosody, will be the focus of this work.
%应该换一种说法，  引出上下文配音即可

To this end, in this work, we propose a Multimodal Context-aware video Dubbing model, termed \textbf{MCDubber}, to convert the modeling object from a single sentence to a longer sequence with context information to ensure the consistency of the global context prosody.
Specifically, MCDubber consists of three key components: 1) a Context Duration Aligner (CDA) incorporates the phonemes and lip frames of the previous, current, and following sentences as inputs to learn context-aware alignment.
% The text-video aligner can also enable generated dubbing to possess audio-visual synchronization.
2) a Context Prosody Predictor (CPP) seeks to read the global context visual sequence and predict the context-aware global energy and pitch. 3) a Context Acoustic Decoder (CAD) predicts the global context mel-spectrogram, assisted by the ground-truth mel-spectrograms of the adjacent sentences. 
Finally, the extracted mel-spectrogram of the current sentence from the output global context mel-spectrograms is converted into time-domain waveforms by the HiFiGAN \cite{hifigan} vocoder, as the final required dubbing audio.
The main contributions of this paper are as follows:

1) We propose MCDubber, a novel AVD model that converts the modeling object from a single sentence to a longer sequence with multimodal context information to ensure the consistency of the global context prosody. To the best of our knowledge, this paper is the first to systematically integrate multimodal context information to enhance prosody expressiveness in AVD.

2) We design a CDA to learn the context-aware alignment between the text and lip frames. Additionally, we proposed a CPP that seeks to read the global context visual sequence and predict the context-aware global energy and pitch. A CAD is designed to ultimately predict the global context mel-spectrogram with the assistance of ground-truth mel-spectrograms of the current sentence. 

3) Extensive experimental results demonstrate that our model significantly improves prosody expressiveness in dubbing and ensures that the generated dubbing aligns with the multimodal context. We further conduct experiments to find the best multimodal context length for expressive video dubbing.
\vspace{-3.5mm}
\section{Related Work}
\vspace{-1.5mm}

In recent years, numerous TTS methods \cite{tacotron1, tacotron2, fastspeech2, glow_tts, natural_speech,  valle} have been developed to generate human-like voice from text. FastSpeech1 \cite{fastspeech1} significantly accelerates mel-spectrogram generation by introducing a Feed-Forward Transformer (FFT). Flow-TTS \cite{flow_tts}, based on generative flow, achieves high-quality speech generation. NaturalSpeech2 \cite{natural_speech2} synthesizes highly expressive, robust, high-fidelity speech using a neural audio codec with continuous latent vectors and a latent diffusion model. Despite their ability to generate high-quality speech, these TTS models are unsuitable for the AVD task due to their lack of facial visual modeling and the need for audio-visual synchronization.

A growing number of methods \cite{masked_speech, context_speech, paratts, oplustil2020using, liu2024emotion, xu2021improving, cong21b_interspeech} in TTS focus on modeling context information to enhance the performance of synthesized speech. ContextSpeech \cite{xiao23_interspeech} uses a memory-cached recurrence mechanism to incorporate global textual and acoustic context into sentence encoding, significantly improving voice quality and prosody expressiveness. Inspired by the masking strategy in speech editing research, MaskedSpeech \cite{masked_speech} incorporates both contextual semantic and acoustic features to enhance prosody generation. Unlike traditional TTS models that only consider textual and acoustic context, our approach integrates multimodal context to enhance prosody expressiveness in dubbing and ensure that the generated prosody aligns with the original video's multimodal context.

\vspace{-3.5mm}
\section{Proposed Method}
\vspace{-1.5mm}

\subsection{Overview}
\begin{figure}[!h]
\centering
\includegraphics[width=\linewidth]{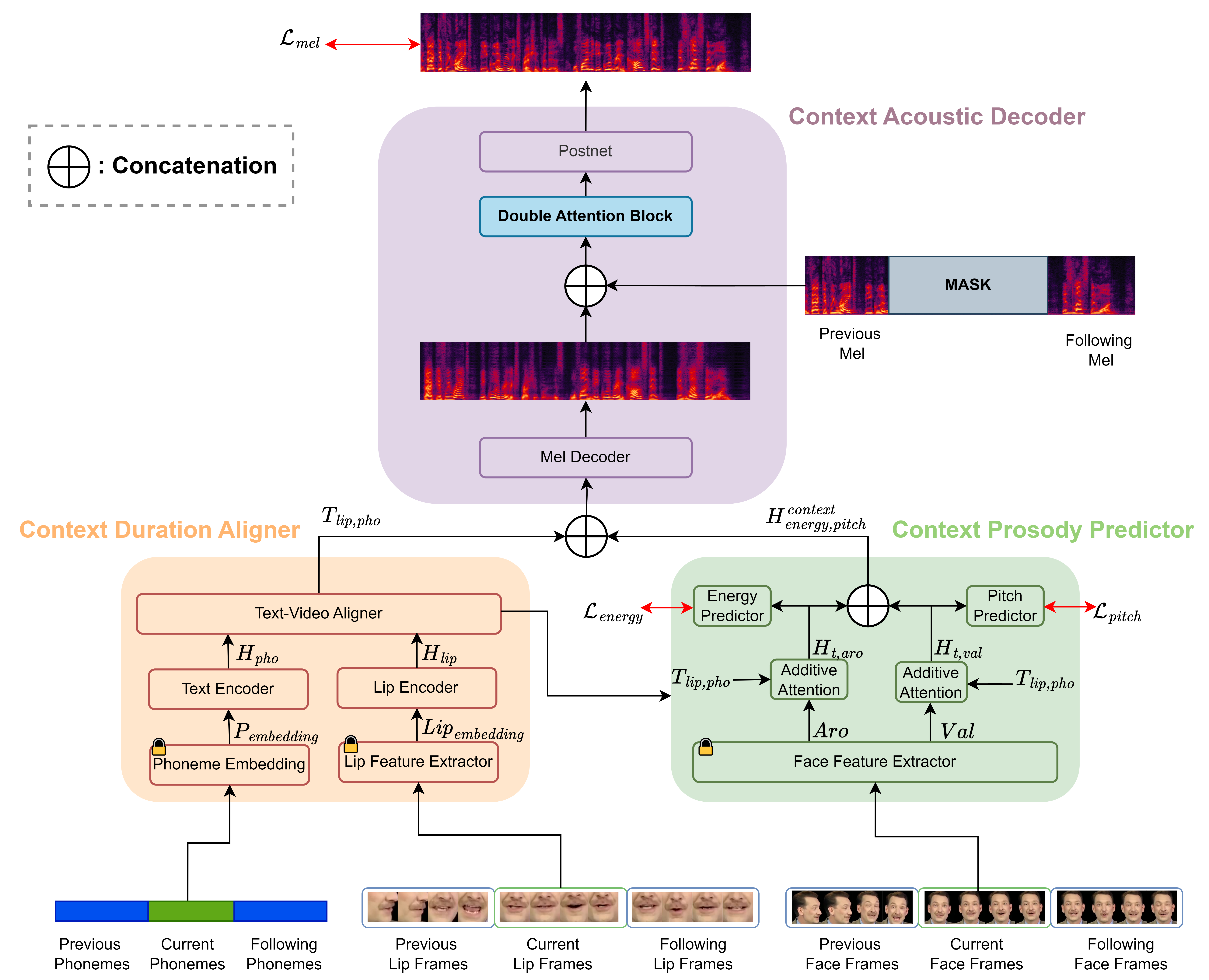}
\caption{The overview of the proposed MCDubber, that consists of the Context Duration Aligner (CDA) (Sec 3.3), the Context Prosody Predictor (CPP) (Sec 3.4), and the Context Acoustic Decoder (CAD) (Sec 3.5). Note that $\oplus$ denotes the concatenation operator.}
\label{fig_model}
\end{figure}

\vspace{-1.5mm}
The proposed MCDubber employs HPMDubbing \cite{HPMDubbing} as the network backbone. It aims to enhance prosody expressiveness in dubbing and ensure that the generated prosody aligns with the original video's multimodal context by extending the modeling from single sentences to longer sequences and integrating multimodal context information. The main architecture of the model is shown in Figure \ref{fig_model}. Unlike existing expressive dubbing models, our method enhances the prosody expressiveness of generated speech while ensuring alignment with the multimodal context by incorporating the multimodal context of the current sentence. The phonemes are first extracted from the raw text $\{ T_{\text{previous}}, T_{\text{current}}, T_{\text{following}} \}$ of the previous, current, and following sentences, while the lip frame sequences are cropped from the video frame sequence $V=\{ V_{\text{previous}}, V_{\text{current}}, V_{\text{following}} \}$. Phonemes and lip frames are then fed into our CDA, which learns context-aware alignment between them. A Text-Video Aligner in CDA is utilized to address the challenges of duration control and audio-visual synchronization. Next, taking the face frames extracted from $V$ and the output of the CDA as input, the CPP enhances dubbing expressiveness by predicting the context-aware global energy and pitch.
The outputs of the CDA and CPP are concatenated and fed to the CAD module, which predicts the global context mel-spectrogram using adjacent ground-truth mel-spectrograms of the current sentence from $\{ A_{\text{previous}},  A_{\text{following}} \}$. Detailed descriptions of each component in our model are provided below.

\vspace{-3.5mm}
\subsection{Preliminaries}
\vspace{-1.5mm}
\subsubsection{Audio-Visual synchronization.} One goal of AVD is to synthesize speech that aligns with lip motion in video. Notably, a natural temporal correspondence exists between audio and video. Specifically, in an audio-visual clip, there is a relationship between the length of a mel-spectrogram sequence and the number of video frames \cite{hu2021neural}:
\begin{equation}
    n = \frac{T_{\mathit{mel}}}{T_v} = \frac{{sr}/{hs}} {\mathit{FPS}} \in \mathbb{N}^+
\end{equation}
where $sr$ denotes the sampling rate of the audio, $FPS$ is the Frames per Second of the video and $hs$ is the hop size of the mel-spectrogram. Audio-visual synchronization can be achieved by aligning text with the lip motion, which determines the speech content.

\vspace{-3.5mm}
\subsubsection{Multimodal Context  Selection.} Considering that excessively long phoneme inputs increase the difficulty of dubbing and that distant multimodal context may not be relevant for predicting the prosody of the current sentence \cite{context_speech}, we set a hyperparameter $K$, referred to as the maximum multimodal context length. Specifically, given the phoneme sequences $P_{\mathit{pre}} = \{P_1, P_2, \ldots, P_{T_{\mathit{pre}}}\}$ and $P_{\mathit{fol}} = \{P_1, P_2, \ldots, P_{T_{\mathit{fol}}}\}$, representing the phonemes of the previous and following sentences, with respective lengths $T_{\mathit{pre}}$ and $T_{\mathit{fol}}$. The $K$ phonemes of each sentence are selected as input to our model: $\{P_{T_{\mathit{pre}} - K + 1}, P_{T_{\mathit{pre}} - K + 2}, \ldots, P_{T_{\mathit{pre}}}\}$ for the previous sentence and $\{P_1, P_2, \ldots, P_K\}$ for the following sentence. Sentences with phoneme sequences length shorter than \(K\) will be entirely selected. The mel-spectrograms corresponding to the selected phonemes are then obtained using the Montreal Forced Aligner \cite{MFA}. According to Equation (2), the video frames corresponding to the selected mel-spectrograms can also be obtained. We will analyze the impact of different $K$ values on our model's enhancement of prosody expressiveness.

\vspace{-4.0mm}
\subsection{Context Duration Aligner}
\vspace{-0.5mm}
The Context Duration Aligner involves two steps : (1) extracting textual representation from the concatenated phoneme sequence and lip representation from the concatenated video frame sequence, and (2) aligning the textual representation with the lip representation.

\vspace{-5.0mm}
\subsubsection{Extracting textual and lip representations.} Let $P_{\mathit{pre}}^{\mathit{selected}}$ and $P_{\mathit{fol}}^{\mathit{selected}}$ represent the selected phoneme sequences of the previous and following sentences, respectively. $P_{\mathit{cur}}^{\mathit{all}}$ represents all phoneme sequences of the current sentence.
A phoneme encoder extracts phoneme embeddings $P_\mathit{embedding}$ from the concatenated phoneme sequence $\{P_{\mathit{pre}}^{\mathit{selected}}, P_{\mathit{cur}}^{\mathit{all}}, P_{\mathit{fol}}^{\mathit{selected}}\}$, and the TextEncoder is utilized to obtain the textual representation:
\begin{equation}
H_{\mathit{pho}} = \text{TextEncoder}(P_\mathit{embedding}) \in \mathbb{R}^{T_p \times D}
\end{equation}
where $T_p$ is the length of $P_\mathit{embedding}$, and the TextEncoder comprises a series of FFT blocks.

Let $V_{\mathit{pre}}^{\mathit{selected}}$ and $V_{\mathit{fol}}^{\mathit{selected}}$ denote the selected video frame sequences of the previous and following sentences, respectively, while $V_{\mathit{cur}}^{\mathit{all}}$ represents all video frame sequences of the current sentence. First, a lip feature extractor \cite{martinez2020lipreading} is employed to extract lip embeddings $\mathit{Lip}_{\mathit{embedding}} \in \mathbb{R}^{T_v \times D}$ from $\{V_{\mathit{pre}}^{\mathit{selected}}, V_{\mathit{cur}}^{\mathit{all}}, V_{\mathit{fol}}^{\mathit{selected}}\}$. Next, we utilize the LipEncoder to obtain lip representations:
\begin{equation}
H_{\mathit{lip}} = \text{LipEncoder}(\mathit{Lip}_{\mathit{embedding}}) \in \mathbb{R}^{T_v \times D}
\end{equation}
where $T_v$ is the length of concatenated video frame sequence, and the LipEncoder comprises a series of FFT blocks.
\vspace{-5.0mm}
\subsubsection{Aligning text with lips.} We employ a Text-Video Aligner \cite{HPMDubbing} to align textual and lip representations using a cross-modal attention mechanism for learning context-aware alignment:
\begin{equation}
\begin{aligned}
    H_{\mathit{lip,pho}} &= \text{Softmax}\left( \frac{H_{\mathit{lip}} H_{\mathit{pho}}^T}{\sqrt{d}} \right) H_{\mathit{pho}} \\
    &= AH_{\mathit{pho}}  \in \mathbb{R}^{T_v \times D}
\end{aligned}
\end{equation}
where $A \in \mathbb{R}^{T_v \times T_p}$. Then transposed convolutions is used to expand $H_{\mathit{lip,pho}}$, which can be formulated as:
\begin{equation}
\begin{aligned}
    T_{\mathit{lip,pho}} &= \text{Conv-Transpose}\left(n, H_{\mathit{lip,pho}}\right)\in \mathbb{R}^{T_{\mathit{mel}}  \times D}
\end{aligned}
\end{equation}
where \( T_{\mathit{mel}} \) denotes the length of the desired mel-spectrogram corresponding to the previous, current, and following sentences. Different existing dubbing models learn alignment for a single sentence, while our model extends the alignment object to a longer sequence, focusing on learning context-aware alignment.

\vspace{-4.0mm}
\subsection{Context Prosody Predictor} 
\vspace{-0.5mm}
One goal of AVD is to generate speech that aligns with prosody expressiveness. Existing video dubbing methods enhance prosody expressiveness by utilizing facial visual information of a single sentence. Unlike previous work, the Context Prosody Predictor analyzes the global context visual sequence and predicts context-aware global energy and pitch. The Context Prosody Predictor contains two steps: (1) extracting context arousal and valence features from concatenated video frame sequence, and (2) predicting context-aware global energy and pitch.
\vspace{-8.0mm}
\subsubsection{Extracting context arousal and valence.} 
According to findings in \cite{HPMDubbing}, arousal and valence information from facial expressions can assist in predicting the energy and pitch of desired speech. An emotion face-alignment network \cite{emofan} is employed to derive context arousal feature $\mathit{Aro} \in \mathbb{R}^{T_v \times D}$ and valence feature $\mathit{Val} \in \mathbb{R}^{T_v \times D}$ from $\mathit{Face}$ which croped from $\{V_{\mathit{pre}}^{\mathit{selected}}, V_{\mathit{cur}}^{\mathit{all}}, V_{\mathit{fol}}^{\mathit{selected}}\}$ using $S^3 FD$ \cite{s3fd}.
\subsubsection{Predicting context-aware global energy and pitch.} First, a context textual arousal-related representation $\mathit{H}_{\mathit{t, aro}}$ is generated by combining the context arousal $\mathit{Aro}$ and the lip-phoneme feature $T_{\mathit{lip, pho}}$ using additive attention:
\begin{equation}
    \begin{aligned} H_{\mathit{t, aro}}^\mathit{i} & =\sum_{k=0}^{T_\mathit{mel}-1} \alpha_{i, k} T_{lip, pho}^k, \\ \alpha_{i, k} & =\exp \left(\hat{\alpha}_{i, k}\right) / \sum_{j=0}^{T_\mathit{mel}-1} \exp \left(\hat{\alpha}_{i, j}\right), \\ \hat{\alpha}_{i, k} & =\mathbf{w}_{\mathbf{a}}^{\top} \tanh \left(\mathbf{W}_{\mathbf{a}}^{\top} \mathit{Aro}_{\mathit{i}}+\mathbf{U}_{\mathbf{a}}^{\top} T_{lip, pho}^k+\mathbf{b}_{\mathbf{a}}\right)\end{aligned}
\end{equation}
where $i$ is frame index, $\mathit{Aro}_{\mathit{i}}$ is the $i$-th row of $\mathit{Aro}$, $\alpha_{i, k}$  is the attention weight on the $k$-th lip-phonme feature $T_{lip, pho}^k$ regarding $i$-th arousal display; Then, context textual arousal-related  representation $\mathit{H}_{\mathit{t, aro}}$ is used to predict context-aware global energy $\hat{E}_{\mathit {pred}}$ via an energy predictor \cite{fastspeech2}. 
% \begin{equation}
% \hat{E}_{\mathit {pred}}=\text { EnergyPredictor }\left(\left\{A_i^l\right\}_{i=1}^{T_\mathit{mel}}\right),
% \end{equation}

The process for predicting context-aware global pitch is similar. First, a context textual valence-related representation $\mathit{H}_{\mathit{t, val}}$ is generated by combining the context valence $\mathit{Val}$ and the lip-phoneme feature $T_{\mathit{lip, pho}}$ using additive attention. Then, context textual valence-related representation $\mathit{H}_{\mathit{t, val}}$ is used to predict context-aware global pitch $\hat{P}_{\mathit {pred}}$ via a pitch predictor\cite{fastspeech2}.
Unlike previous work, we extend prosody prediction from a single sentence to longer sequences using multimodal context, predicting context-aware global prosody to effectively enhance the expressiveness of dubbing. Finally, we concatenate the context-aware prosody-related global features $\mathit{H}_{\mathit{t, aro}}$ and $\mathit{H}_{\mathit{t, val}}$ as the output $H_{\mathit{energy,pitch}}^ \mathit{context}$ of CPP.
\vspace{-4.0mm}
\subsection{Context Acoustic Decoder}
\vspace{-0.5mm}
In continuous videos, human speech maintains consistency, with the prosody of the current sentence influenced by context speech \cite{masked_speech}. In the AVD task, the generated dubbing will be combined with the original context in the final video. Therefore, ensuring that the prosody of the generated dubbing aligns with the multimodal context is crucial, necessitating the consideration of context ground-truth speech. Therefore, We designed a Context Acoustic Decoder to predict global context mel-spectrograms with the assistance of adjacent ground-truth mel-spectrograms of the current sentence.

First, we concatenate the outputs $T_{\mathit{lip, pho}}$ from the CDA and $H_{\mathit{energy, pitch}}^{\mathit{context}}$ from the CPP.  The concatenated result is then converted into a mel-spectrogram sequence $\hat{F}_{\mathit{mel}}$ using a Mel Decoder \cite{fastspeech2}. The mel-spectrogram of the current sentence is masked and combined with the adjacent ground-truth mel-spectrograms to form $M = \{M_{\mathit{pre}}^{\mathit{selected}}, \mathit{MASK}, M_{\mathit{fol}}^{\mathit{selected}}\}$. $M$ is then concatenated with \(\hat{F}_{\mathit{mel}}\) along the hidden dimension and used as input for a Double Attention Block (DAB) \cite{double_attention_block}. The role of DAB is to aggregate and propagate global features from the concatenation of \(\hat{F}_{\mathit{mel}}\) and $M$. Specifically, the DAB collects features through attention pooling and selects and distributes them to each location via an attention mechanism. Finally, a Postnet \cite{tacotron2} is used to ultimately predict global context mel-spectrogram $\hat{Y}_{\mathit{mel}}$.
The extracted mel-spectrogram of the current sentence from the global context mel-spectrograms is converted into a time-domain waveform using HiFi-GAN \cite{hifigan} as the final required dubbing audio. Through the above process, considering adjacent ground-truth speech, the CAD ensures that the prosody of the dubbing audio aligns with the multimodal context.

\vspace{-4.0mm}
\subsection{Loss Functions}
\vspace{-0.5mm}
Our proposed model is optimized using the following loss function:

% \begin{equation}
% \small
% \mathcal{L}_{\mathit {energy }} =\frac{1}{T_\mathit{mel}} \sum_{t=0}^{T_\mathit{mel}-1}\left(E_\mathit{gt}^{\mathit{t}}-\hat{E}_{\mathit {pred}} ^{\mathit{t}}\right)^2, \mathcal{L}_{\mathit {pitch }} =\frac{1}{T_\mathit{mel}} \sum_{t=0}^{T_\mathit{mel}-1}\left(P_\mathit{gt}^{\mathit{t}}-\hat{P}_{\mathit {pred}} ^{\mathit{t}}\right)^2 
% \end{equation}

% \begin{equation}
% \small
% \mathcal{L}_{\mathit {mel }}^{f} =\frac{1}{T_\mathit{mel}} \sum_{t=0}^{T_\mathit{mel}-1} ||Mel_\mathit{gt}^{\mathit{t}}-\hat{F}_{\mathit {pred}}^{\mathit{t}}|| ,\mathcal{L}_{\mathit {mel }}^{y} =\frac{1}{T_\mathit{mel}} \sum_{t=0}^{T_\mathit{mel}-1} ||Mel_\mathit{gt}^{\mathit{t}}-\hat{Y}_{\mathit {pred}}^{\mathit{t}}||
% \end{equation}

\begin{equation}
\mathcal{L}_{\mathit {sum }} = \mathcal{L}_{\mathit {energy }} + \mathcal{L}_{\mathit {pitch }} + \mathcal{L}_{\mathit {mel }} 
\end{equation}
Where Mean Squared Error (MSE) is used as the training objective for $\mathcal{L}_{\mathit{energy}}$ and $\mathcal{L}_{\mathit{pitch}}$, while Mean Absolute Error (MAE) is used for $\mathcal{L}_{\mathit{mel}}$. We compute these training objectives for the previous, current, and following sentences.

% \begin{equation}
% \mathcal{L}_{\mathit {sum }} = \mathcal{L}_{\mathit {energy }} + \mathcal{L}_{\mathit {pitch }} + \mathcal{L}_{\mathit {mel }}^{f} + \mathcal{L}_{\mathit {mel }}^{y} + \mathcal{L}_{\mathit {ctc}}
% \end{equation}

% CTC loss $\mathcal{L}_{\mathit {ctc}}$ is used to shape diagonal attention. 

% \subsection{Training}

\vspace{-3.5mm}
\section{Experiments}
\vspace{-1.5mm}
Datasets with consecutive samples are relatively scarce, so we evaluate our model solely on the Chem dataset. We first introduce the Chem dataset, followed by implementation details and training strategy, and finally, present the evaluation metrics.
\vspace{-4.0mm}
\subsection{Dataset}
\vspace{-1.0mm}
The Chem dataset is a single-speaker English speech dataset consisting of 6,319 short video clips and corresponding transcripts collected from YouTube \cite{chemdataset}, featuring rich speech prosody. Initially, 5,938 samples were assigned to the training set, 187 to the validation set, and 194 to the test set. During the collection process, clips lacking the speaker's face were filtered out, resulting in the presence of non-consecutive samples in the dataset. Consequently, we re-collected consecutive samples, which include the previous, current, and following sentences, with the current sentence being the target for dubbing. This re-collected dataset is referred to as the Context Chem dataset. The new training set has 3,308 consecutive samples, the new validation set has 85, and the new test set has 113. Therefore, the consecutive samples in the new test set will be used to evaluate our model.

% \begin{table}[h!]
% \setlength{\abovecaptionskip}{10pt} 
% \setlength{\belowcaptionskip}{-10pt}
% \caption{Sample Statistics of the Chem Dataset}
% \label{table1}
% \centering
% \begin{tabular}{p{20mm}<{\centering} p{40mm}<{\centering} p{30mm}<{\centering}}
% \toprule
% {Set} & Total Samples & Consecutive Samples \\ \hline
% Train & 5938 & 3308 \\ \hline
% Validation & 187 & 85 \\ \hline
% Test & 194 & 113 \\ \bottomrule
% \end{tabular}
% \end{table}
\vspace{-4.0mm}
\subsection{Implementation Details}
\vspace{-1.0mm}

For the CDA, we use 6 FFT blocks for both the TextEncoder and LipEncoder, with the dimensions of phoneme representation  \( H_{\mathit{pho}} \) and lip representation  \( H_{\mathit{lip}} \) set to 256. We employ 8 attention heads for the Text-Video Aligner. Each video clip is processed at 25 FPS with a sampling rate of 16000 Hz. A short-time Fourier transform (STFT) with 1024 points is used to obtain the mel-spectrogram. In the CPP, the dimensions of \( \mathit{H}_{\mathit{t, aro}} \) and \( \mathit{H}_{\mathit{t, val}} \) are set to 256. The CAD's Mel Decoder consists of 6 FFT blocks.

For training, we employed the Adam optimizer with an initial learning rate set to 0.0002, \( \beta_1 = 0.9 \), \( \beta_2 = 0.98 \), and \( \epsilon = 10^{-9} \). The batch size is set to 8. Our model, implemented in PyTorch, is trained on a single NVIDIA A100 80G GPU. The maximum multi-modal context length \( K \) is set to 50. The configuration of the hyperparameter \( K \) will be thoroughly discussed in Section 5.1.

\vspace{-3.0mm}
\subsection{Training Strategy}
\vspace{-0.5mm}

Due to the limited number of samples in the Context Chem Dataset compared to the original Chem dataset, the model struggles with effective training when data is scarce. To address this issue, our training process is divided into two stages: 
\textbf{Stage 1}: Train the model for 800 epochs using the total samples from the training set in the original Chem dataset under a single sentence dubbing conditions, following the traditional dubbing model training process \cite{HPMDubbing}. \textbf{Stage 2}: Initialize the model with the checkpoint from \textbf{Stage 1} and continue training using samples from the training set in the Context Chem dataset, incorporating multimodal contextual information.
This training strategy not only addresses the issue of insufficient training data but also enhances the overall performance of our model. The effectiveness of this training strategy will be examined in Section 5.3.

\vspace{-3.0mm}
\subsection{Evaluation Metrics}
\vspace{-0.5mm}

\subsubsection{Objective metrics.}
In the objective evaluations, we used several metrics to evaluate the
synthesized speech:
(1) Gross Pitch Error (GPE) \cite{GPE}: measures the percentage of frames where the pitch error exceeds 20\% and voicing is present in both the synthesized speech and ground-truth speech. 
(2)F0 Frame Error (FFE) \cite{FFE}: measures the percentage of frames with either a voicing decision error or a pitch error exceeding 20\%. The GPE and FFE metrics are related to prosody; lower values indicate that the synthesized speech demonstrates greater prosody expressiveness. 
(3) Short-Time Objective Intelligibility (STOI) \cite{stoi}: estimate the speech intelligibility. 
(4) Lip Sync Error-Confidence (LSE-C) \cite{syncnet, lipexpert}: represents the average confidence score. Higher scores indicate better audio-video correlation.
(5) Lip Sync Error-Distance (LSE-D) \cite{syncnet, lipexpert}: represents the average error by measuring the distance between lip and audio representations. A lower LSE-D indicates better audio-visual synchronization.

\vspace{-0.5mm}
\subsubsection{Subjective metrics.}

We conducted a Mean Opinion Score (MOS) test with 20 raters, all of whom received specialized training on the evaluation criteria. The raters were asked to evaluate 12 generated dubbing and speech samples, rating the following metrics on a scale from 1 to 5: (1) Audio Quality(AQ) and Audio-Visual Synchronization (AV Syn): overlay the synthesized speech onto the original video and assess both AQ and AV Syn.
(2) MOS-Context (MOS-C): After dubbing the current sentence, splice the video of dubbing with the contextual original video to evaluate how well the prosody expressiveness of the generated dubbing aligns with the multimodal context.
(3) MOS-Similarity (MOS-S): Evaluate the similarity of the synthesized speech to the ground truth in terms of prosody.
(4) MOS-Naturalness (MOS-N): Evaluate the naturalness of the synthesized speech.

\vspace{-3.5mm}
\section{Results and Disscussion}

\subsection{Multimodal Context Length Analysis}

To determine a suitable range for the hyperparameter $K$, which represents the maximum multimodal context length, we performed a statistical analysis on the phoneme sequence length for the previous and following sentences within the Context Chem dataset. As demonstrated in Table \ref{table2}, the mean length of phoneme sequence in the previous and following sentences is approximately 46.
Therefore, we trained our model with $K$ values of $\{10, 20, 30, 40, 50, 60\}$ to study the prosody expressiveness of the generated speech. The results are presented in Table \ref{table3}. All models perform similarly in terms of STOI, as expected due to the proposed method's primary focus on prosody expressiveness. When \( K = 50 \), the speech generated by our model exhibits the lowest values of prosody-related metrics GPE and FFE. Lower values of GPE and FFE indicate that the prosody of the synthesized speech is closer to that of the ground-truth speech, demonstrating greater prosody expressiveness. This suggests that appropriately increasing the maximum multimodal context length can enhance the prosody expressiveness of synthesized speech. However, with increasing $K$, the LSE-C, and LSE-D metrics deteriorate, implying that excessively long phoneme sequences input increases dubbing difficulty. Hence, we select $K=50$ as the optimal maximum multimodal context length to strike a balance between prosody expressiveness and dubbing difficulty.
\begin{table}[h!]
\setlength{\abovecaptionskip}{3pt} 
\setlength{\belowcaptionskip}{-3pt}
\caption{Statistics results of phoneme sequence length for Previous and Following sentences within the Context Chem dataset.}
\label{table2}
\centering
\resizebox{0.65\textwidth}{!}{
\begin{tabular}{p{15mm}<{\centering} p{5mm}<{\centering} p{15mm}<{\centering} p{15mm}<{\centering} p{15mm}<{\centering} p{15mm}<{\centering}}
\toprule
Context &&    & Training & Validation & Test \\ \hline
\multirow{4}{*}{Previous} &&  Mean & 46 & 46 & 48 \\
 &&  Median & 43 & 43 & 47 \\
 &&  Min & 2 & 2 & 7 \\
 &&  Max & 133 & 133 & 119 \\ \hline
\multirow{4}{*}{Following} &&  Mean & 45 & 45 & 45 \\
 && Median & 42 & 42 & 43 \\
 && Min & 2 & 2 & 6 \\
 && Max & 132 & 132 & 119 \\ \bottomrule
\end{tabular}
}
\end{table}

\begin{table}[ht]
\setlength{\abovecaptionskip}{3pt} 
\setlength{\belowcaptionskip}{-3pt}
\caption{The results of the objective evaluation for proposed MCDubber with different max multimodal context length $K$. ↑(↓) indicates that a higher (lower) value is better, and bold indicates the best score.}
\label{table3}
\centering
\renewcommand{\arraystretch}{1.2} % Adjust the row height
\resizebox{0.55\textwidth}{!}{% Resize the table to fit text width

\begin{tabular}{cccccccccccccc}
\hline
& & $K$ \qquad & & GPE ↓ \qquad & & FFE ↓ \qquad & & STOI ↑ \qquad & & LSE-C ↑ \qquad & & LSE-D ↓ \qquad & \\ % Adjusted headers, removed WER column
\hline
& & 10 \qquad & & 41.89 \qquad & & 33.20 \qquad & & 0.507 \qquad & & \textbf{7.999} \qquad & & \textbf{6.891} \qquad & \\
& & 20 \qquad & & 40.82 \qquad & & 32.59 \qquad & & 0.501 \qquad & & 7.729 \qquad & & 6.911 \qquad & \\
& & 30 \qquad & & 42.00 \qquad & & 32.35 \qquad & & 0.490 \qquad & & 7.700 \qquad & & 6.945 \qquad & \\
& & 40 \qquad & & 44.08 \qquad & & 31.13 \qquad & & 0.499 \qquad & & 7.644 \qquad & & 6.922 \qquad & \\
& & 50 \qquad & & \textbf{40.26} \qquad & & \textbf{30.94} \qquad & & \textbf{0.508} \qquad & & 7.571 \qquad & & 6.943 \qquad & \\
& & 60 \qquad & & 41.43 \qquad & & 31.33 \qquad & & 0.507 \qquad & & 7.560 \qquad & & 6.935  \qquad & \\
\hline
\end{tabular}%
}
\label{tab:performance_metrics}
\end{table}

\vspace{-4.0mm}
\subsection{Results of Video Dubbing}
To evaluate the prosody expressiveness of MCDubber, we develop three baseline models as follows. (1) \textbf{FastSpeech2} \cite{fastspeech2}: a neural TTS model; (2) \textbf{DSU-AVO} \cite{DSU-AVO}, an AVD model with a learning objective focused on discrete speech unit prediction; (3) \textbf{HPMDubbing} \cite{HPMDubbing}, an AVD model incorporating hierarchical prosody modeling considering facial expressions. For a fair comparison, these baseline models were trained on the Original Chem dataset. Note that all of the above baseline models are designed for modeling a single sentence.
The results, shown in Table \ref{table4_evaluations}, indicate that our method achieves the best performance on prosody-related metrics. Specifically, our method attains a GPE of 40.26 and an FFE of 30.94, surpassing previous methods. This demonstrates that our model generates more prosody expressive speech. Additionally, our method also achieved the best performance in prosody-related metrics, specifically MOS-S and MOS-C, as determined by human subjective evaluations. MOS-S shows our synthesized speech is closer to ground-truth speech in terms of prosody. MOS-C indicates that after integrating the dubbed video with the original context, the prosody of the generated dubbing aligns well with the multimodal context, outperforming other models. In summary, both objective and subjective metrics demonstrate that our model improves dubbing prosody expressiveness and enhances the prosody consistency of the dubbed video with the multimodal context compared to other methods.

\begin{table}[ht]
\setlength{\abovecaptionskip}{3pt} 
\setlength{\belowcaptionskip}{-3pt}
\caption{The results of objective and subjective evaluations with comparisons against other methods. ↑(↓) indicates that a higher (lower) value is better, and bold indicates the best score.}
\label{table4_evaluations}
\centering
\renewcommand{\arraystretch}{1.2} % Adjust the row height
\resizebox{1.0\textwidth}{!}{%
\begin{tabular}{ccccccccccc}
\hline
Methods & GPE ↓ & FFE ↓ & STOI ↑ & LSE-C ↑ & LSE-D ↓ & AQ ↑  & AV Sync ↑ & MOS-C ↑ & MOS-S ↑ & MOS-N ↑ \\
\hline
GT & N/A & N/A & N/A & 8.722 & 6.564 & 4.23 ± 0.08 & 4.39 ± 0.07 & 4.41 ± 0.07 & 4.44 ± 0.07 & 4.46 ± 0.07 \\
Mel Resynthesis  & 02.77 & 09.70 & 0.952 & 8.468 & 6.538 & 4.14 ± 0.05 & 4.20 ± 0.07 & 4.21 ± 0.05 & 4.27 ± 0.07 & 4.24 ± 0.06 \\
\hline
FastSpeech2 & 50.26 & 41.24 & 0.211 & 3.438 & 9.399 & 3.50 ± 0.12 & 3.36 ± 0.08 & 3.53 ± 0.06 & 3.39 ± 0.11 & 3.44 ± 0.12 \\
DSU-AVO & 48.02 & 32.31 & \textbf{0.536} & \textbf{8.198} & 7.120 & 3.77 ± 0.06 & 3.74 ± 0.06 & 3.55 ± 0.06 & 3.72 ± 0.10 & 3.90 ± 0.10 \\
HPMDubbing & 42.78 & 36.56 & 0.383 & 7.934 & 7.057 & 3.69 ± 0.05 & 3.81 ± 0.05 & 3.84 ± 0.07 & 3.82 ± 0.08 & 3.91 ±0.07 \\
\hline
Ours & \textbf{40.26} & \textbf{30.94} & 0.508 & 7.571 & \textbf{6.943} & \textbf{3.83 ± 0.05} & \textbf{3.97 ± 0.06} & \textbf{4.02 ± 0.06} & \textbf{4.03 ± 0.08} & \textbf{3.96 ± 0.07} \\
\hline
\end{tabular}%
}
\end{table}

\begin{table}[ht]
\setlength{\abovecaptionskip}{3pt} 
\setlength{\belowcaptionskip}{-3pt}
\caption{The results of objective and subjective evaluations for various ablation models. ↑(↓) indicates that a higher (lower) value is better, and bold indicates the best score.  PRE, FOL, and TS represent the multi-modal context information of the previous sentence, the following sentence, and the training strategy described in Section 4.3, respectively.}
\label{table_ablation_methods}
\centering
\renewcommand{\arraystretch}{1.2} % Adjust the row height
\resizebox{0.75\textwidth}{!}{%
\begin{tabular}{cccccccccc}
\hline
Methods & GPE ↓ & FFE ↓ & STOI ↑ & LSE-C ↑ & LSE-D ↓ & MOS-C ↑ & MOS-S ↑ \\ 
\hline
w/o CDA & 42.53 & 34.27 & 0.501 & 8.066 & 6.901 & 3.82 ± 0.08 & 3.80 ± 0.11  \\
w/o CPP & 40.57 & 32.37 & \textbf{0.523} & 8.089 & 6.980 & 3.77 ± 0.09 & 3.81 ± 0.10  \\
w/o CAD & 40.64 & 32.73 & 0.509 & \textbf{8.098} & 6.893 & 3.71 ± 0.06 & 3.89 ± 0.10 \\
\hline
w/o PRE & 40.87 & 31.81 & 0.512 & 7.846 & 6.897 & 3.84 ± 0.08 & 3.76 ± 0.12  \\
w/o FOL & 41.38 & 32.57 & 0.498 & 7.925 & \textbf{6.865} & 3.87 ± 0.09 & 3.82 ± 0.10  \\
\hline
w/o TS & 58.41 & 45.80 & 0.206 & 1.956 & 10.043 & 1.94 ± 0.10 & 1.92 ± 0.07  \\
\hline
Ours & \textbf{40.26} & \textbf{30.94} & 0.508 & 7.571 & 6.943 & \textbf{4.02 ± 0.06} & \textbf{4.03 ± 0.08}  \\
\hline
\end{tabular}%
}
\end{table}
\vspace{-4.0mm}
\subsection{Ablation Studies}

To investigate the effects of each component in MCDubber, the multimodal information from adjacent sentences, and the training strategy, we conducted a comprehensive ablation study.
\vspace{-4.0mm}
\subsubsection{Effectiveness of Context Duration Aligner, Context Prosody Predictor, Context Acoustic Decoder.} To evaluate the effectiveness of the CDA, CPP, and CAD components of our model in enhancing prosody expressiveness and maintaining multimodal contextual coherence in dubbed videos, we conducted an ablation study by individually excluding the texts, facial video frames, and mel-spectrograms of both previous and following sentences. The detailed results are presented in Rows 2-4 of Table \ref{table_ablation_methods}. Our findings underscore the significant contributions of all proposed components to prosody expressiveness and multimodal contextual coherence. Specifically, removing the CDA results in a significant decrease in the subjective context prosody-related metric (MOS-C), highlighting the importance of the CDA in aligning the dubbed video with the multimodal context. Additionally, excluding the CPP leads to decreased performance metrics related to prosody (FFE, GPE, MOS-S), indicating that the context-aware global prosody learned by CPP enhances the prosody expressiveness of our synthesized speech. Similarly, omitting the CAD results in diminished performance metrics (FFE, GPE, MOS-S, MOS-C), emphasizing CAD's importance in facilitating the generation of prosody expressive and contextually coherent dubbed videos by predicting the global context mel-spectrogram.
\vspace{-4.0mm}
\subsubsection{Effectiveness of Multimodal Context information from the previous and following sentence.} In our model, we integrate multimodal context information from both previous and following sentences to enhance prosody expressiveness and align prosody with the multimodal context. To evaluate their effectiveness, we conducted an ablation study by individually removing the multimodal information input from the previous and following sentences. As shown in Rows 5-6 of Table \ref{table_ablation_methods}, both are crucial for enhancing prosody expressiveness and maintaining multimodal context consistency. Removing multimodal context input information from the previous and following sentences degrades the model's performance in objective metrics related to prosody (FFE, GPE), underscoring its effectiveness in enhancing prosody expressiveness in synthesized speech. Furthermore, prosody-related subjective human evaluation metrics (MOS-S and MOS-C) also decline, suggesting that multimodal context information in adjacent sentences significantly improves the prosody expressiveness and multimodal context consistency of the dubbed video.
\vspace{-4.0mm}
\subsubsection{Effectiveness of Training Strategy.} 
To evaluate the impact of the training strategy on model performance, we trained our model only using consecutive samples in the training set from the Context Chem dataset. Row 7 of Table \ref{table_ablation_methods} shows that the model achieved very poor performance across all metrics. This demonstrates the effectiveness and necessity of our proposed training strategy.
\vspace{-3.5mm}
\section{Conclusion}
\vspace{-1.5mm}

In this paper, we propose MCDubber for expressive video dubbing, which incorporates multimodal context information to enhance the prosody expressiveness of the synthesized speech and ensure that the prosody of the dubbing aligns with the multimodal context. MCDubber employs a context duration aligner designed to learn context-aware alignment, a context prosody predictor aimed at learning context-aware global prosody, and a context acoustic decoder that ultimately predicts the global context mel-spectrogram. Extensive experiments on the Chem benchmark dataset demonstrate that our MCDubber significantly improves dubbing expressiveness compared to advanced baselines. In future work, we will explore the influence of multimodal context on modeling emotion expressiveness in AVD task.

\normalem % 参考文献去除下划线
\bibliographystyle{unsrt}

\bibliography{ref_unsort}
% \end{thebibliography}
\end{document}